\documentclass[pra,twocolumn]{revtex4}
\usepackage{graphicx}
\usepackage{bm}
\usepackage{bbm}
\usepackage{amsmath}

\begin{document}

\title{Increased Dimensionality of Raman Cooling in a Slightly Nonorthogonal Optical Lattice}

\author{Andreas Neuzner}
\altaffiliation{Present address: OHB System AG, Manfred-Fuchs-Str.\ 1, 82234 We{\ss}ling, Germany}
\author{Stephan D\"urr}
\author{Matthias K\"orber}
\author{Stephan Ritter}
\altaffiliation{Present address: TOPTICA Photonics AG, Lochhamer Schlag 19, 82166 Gr\"afelfing, Germany}
\author{Gerhard Rempe}
\affiliation{Max-Planck-Institut f\"{u}r Quantenoptik, Hans-Kopfermann-Str.\ 1, 85748 Garching, Germany}

\begin{abstract}
We experimentally study the effect of a slight nonorthogonality in a two-dimensional optical lattice onto resolved-sideband Raman cooling. We find that when the trap frequencies of the two lattice directions are equal, the trap frequencies of the combined potential exhibit an avoided crossing and the corresponding eigenmodes are rotated by 45$^\circ$ relative to the lattice beams. Hence, tuning the trap frequencies makes it possible to rotate the eigenmodes such that both eigenmodes have a large projection onto any desired direction in the lattice plane, in particular, onto the direction along which Raman cooling works. Using this, we achieve two-dimensional Raman ground-state cooling in a geometry where this would be impossible, if the eigenmodes were not rotated. Our experiment is performed with a single atom inside an optical resonator but this is inessential and the scheme is expected to work equally well in other situations.
\end{abstract}

\maketitle

\section{Introduction}

Limited optical access can hamper laser cooling of atomic gases. A typical example is an experiment with atoms inside a high-finesse Fabry-Perot resonator, where a short resonator is chosen to minimize the mode volume and thus achieve strong coupling. Such systems offer exciting possibilities in the field of quantum information \cite{Reiserer:15}. To avoid fluctuations in the atom-cavity coupling strength, it is desirable to trap the atom in a tightly confining potential and cool the atomic motion to the ground state. An established technique for this is resolved-sideband Raman cooling, which was first demonstrated in ion traps \cite{Monroe:95:Raman}. Subsequently, Raman cooling of optically-trapped neutral atoms was pioneered in a series of experiments that reached the motional ground state in one \cite{Perrin:98}, two \cite{Hamann:98}, and three dimensions \cite{Kerman:00}.

If a Fabry-Perot resonator is short, this typically restricts the optical access to the cavity axis ($z$ axis) and the $xy$ plane. Moreover, a good fraction of the $xy$ plane is often obstructed by mirror-mounting components. Nevertheless, intra-cavity Raman cooling was achieved in one \cite{Boozer:06} and three dimensions \cite{Reiserer:13:PRL}. A crucial aspect in Raman cooling is that the differential photon recoil in the Raman transition must have a noticeable projection onto the principal axis of motion that one wishes to cool. On top of the optical access along the $x$, $y$ and $z$ axes to establish the three-dimensional (3D) optical lattice, Ref.\ \cite{Reiserer:13:PRL} therefore used additional optical access in the $xy$ plane under angles of $45^\circ$ and $-135^\circ$ relative to the $x$ axis to apply two Raman beams, which were combined with a third Raman beam along the cavity axis. A recent experiment \cite{Reimann:14} on two-dimensional (2D) Raman cooling in a 2D lattice, demonstrated that it suffices to apply Raman light along both lattice axes to achieve 2D cooling. This is nontrivial because of an interference effect that forbids the carrier and certain vibrational sideband transitions in some geometries \cite{Boozer:08}. To circumvent this problem, Ref.\ \cite{Reimann:14} drove second-order sideband transitions that are suppressed by the square of the Lamb-Dicke parameter, which makes these transitions quite weak. Hence, the mean number of vibrational excitations was experimentally limited to $0.3(2)$ along the direction which is difficult to cool.

Here, we demonstrate an alternative and novel technique for intra-cavity 2D Raman ground-state cooling in a 2D lattice. The technique requires optical access only along the lattice axes. It is based on cross-dimensional mixing which is created because the lattice geometry is slightly nonorthogonal and the trap frequencies are degenerate. We present a model of this effect and experimentally verify that the trap eigenfrequencies exhibit an avoided crossing when the trap frequencies of the individual lattice beams are nearly degenerate. In addition, the eigenmodes of the trap are predicted to rotate substantially when varying the ratio of the trap frequencies. We verify this by measuring the areas under the Raman sidebands, which are sensitive to the rotation angle. Based on a rotation of $\approx45^\circ$, we perform 2D Raman cooling, reaching a mean excitation number which is estimated to lie between 0.09 and 0.14 for each of the two lattice directions.

\section{Theory}

\subsection{Cross-Dimensional Mixing}

We consider a 2D optical lattice which is formed by two intersecting standing light waves. We assume that the electric field of the $j$-th standing wave with $j\in\{1,2\}$ is monochromatic $\bm E_j(\bm x,t)= \frac12\bm E_{j,0}(\bm x)e^{-i\omega_{j,0} t}+\text{c.c.}$ with spatial profile $\bm E_{j,0}(\bm x)= \bm E_{j,1}\cos[\bm k_j\cdot(\bm x-\bm x_{j,0})]$. Here, $\bm E_{j,1}$ denotes the amplitude, $\bm k_j$ the wave vector, $\omega_{j,0}$ the angular frequency, and $\bm x_{j,0}$ the position at an antinode. We assume that both standing waves are far detuned from all atomic resonances so that the only effect that one standing wave alone has for an atom at position $\bm x$ is to create an optical dipole potential $V_j(\bm x)= -\frac14\alpha_j|\bm E_{j,0}(\bm x)|^2$, where $\alpha_j$ is the dynamic polarizability at the light frequency $\omega_{j,0}$. We use $V_{j,0}= V_j(\bm x_{j,0})$ to denote the dipole potential at an antinode.

If both standing waves are applied simultaneously, an interference term is created. However, we assume that this term oscillates so rapidly in time that the atomic motion cannot follow it. Hence, the atoms effectively experience only the time-averaged potential $V(\bm x)= V_1(\bm x)+V_2(\bm x)$.

We choose the coordinate origin at a minimum of $V(\bm x)$. Depending on whether the potential of the $j$-th standing wave is attractive ($\alpha_j>0$) or repulsive ($\alpha_j<0$), this yields $\bm k_j\cdot\bm x_{j,0}=0$ or $\bm k_j\cdot\bm x_{j,0}=\pi/2$. Hence, a Taylor expansion near the coordinate origin yields the harmonic approximation
\begin{align}
\label{V-k-j}
V(\bm x)
= V(0) + \frac12 \sum_{j=1}^2 \kappa_j\frac{(\bm k_j\cdot\bm x)^2}{k_j^2} +O(r^4)
\end{align}
with the spring constant $\kappa_j= 2|V_{j,0}|k_j^2$ for the $j$-th standing wave.

Let $\pi/2-\varphi$ denote the angle between the wave vectors $\bm k_1$ and $\bm k_2$. Without loss of generality we choose $0\leq \varphi\leq \pi/2$. To achieve this, we might need to reverse the sign of one of the wave vectors. We choose the coordinate system such that
\begin{subequations}
\label{k-1-2}
\begin{align}
\bm k_1
&= k_1(\cos\varphi\bm e_x+\sin\varphi\bm e_z)
,\\
\bm k_2
&= k_2\bm e_z
,\end{align}
\end{subequations}
where $\bm e_x,\bm e_y,\bm e_z$ are cartesian unit vectors. To simplify the following mathematical treatment, we introduce a coordinate system $(x',y',z')$ which is rotated around the $y$ axis by an angle $\varphi/2$. Here, we obtain the symmetric expression
\begin{subequations}
\label{k-1-2-prime}
\begin{align}
\bm k_1
&= k_1\left(\cos\frac\varphi2\bm e_x' +  \sin\frac\varphi2 \bm e_z'\right)
,\\
\bm k_2
&= k_2\left(\sin\frac\varphi2\bm e_x' + \cos\frac\varphi2\bm e_z'\right)
.\end{align}
\end{subequations}
Combination of Eqs.\ \eqref{V-k-j} and \eqref{k-1-2-prime} yields an expansion of the potential in the coordinates $(x',y',z')$
\begin{align}
V(\bm x')
= V(0) + \frac12 \sum_{i,k\in\{1,3\}} x_i'\kappa_{ik}x_k' + O(r^4)
.\end{align}
The coefficients $\kappa_{ik}$ form the spring-constant tensor
\begin{align}
\label{kappa}
\kappa
=
\frac{\kappa_1+\kappa_2}2[\mathbbm 1+(\sin\varphi)\sigma_x]
+ \frac{\kappa_1-\kappa_2}2 (\cos\varphi) \sigma_z
\end{align}
with the $2\times2$ identity matrix $\mathbbm 1$ and the Pauli matrices
\begin{align}
\sigma_x
=
\begin{pmatrix}
0 & 1 \\ 1 & 0
\end{pmatrix}
,&&
\sigma_z
=
\begin{pmatrix}
1 & 0 \\ 0 & -1
\end{pmatrix}
.\end{align}
Diagonalization of the symmetric tensor $\kappa$ is straightforward and yields eigenvalues
\begin{align}
\label{kappa-pm}
\kappa_{\pm}
= \frac{\kappa_1+\kappa_2}2 \pm \frac12\sqrt{\kappa_1^2+\kappa_2^2-2\kappa_1\kappa_2\cos(2\varphi)}
.\end{align}
The corresponding trap angular frequencies are
\begin{align}
\label{omega-kappa}
\omega_j= \sqrt{\frac{\kappa_j}m}
,&&
\omega_\pm= \sqrt{\frac{\kappa_\pm}m}
,\end{align}
where $m$ is the atomic mass.

The corresponding eigenvectors $\bm e_+$ and $\bm e_-$ define the principal axes of the harmonic potential, along which the eigenmodes of the atomic motion are oriented. They can be written as
\begin{subequations}
\label{e-pm-prime}
\begin{align}
\bm e_+
&= \cos\frac\beta2\bm e_x' + \sin\frac\beta2\bm e_z'
,\\
\bm e_-
&= -\sin\frac\beta2\bm e_x' + \cos\frac\beta2 \bm e_z'
.\end{align}
\end{subequations}
Note that unlike the wave vectors $\bm k_1$ and $\bm k_2$, the principal axes $\bm e_+$ and $\bm e_-$ are always mutually orthogonal. The angle $\beta$ is given by
\begin{align}
\label{arg-beta}
\beta
=\arg [(\kappa_1-\kappa_2) \cos\varphi+i(\kappa_1+\kappa_2)\sin\varphi]
,\end{align}
where $\arg z$ with $-\pi<\arg z\leq\pi$ denotes the argument of a nonzero complex number $z= |z|e^{i\arg z}$. With $\kappa_j\geq0$ and $0\leq \varphi\leq \pi/2$, we obtain $0\leq \beta\leq \pi$. Note that Eq.\ \eqref{arg-beta} implies
\begin{align}
\label{tan-beta}
\tan\beta
= \frac{\kappa_1+\kappa_2}{\kappa_1-\kappa_2} \tan\varphi
.\end{align}

\begin{figure}[!tb]
\includegraphics[width=\columnwidth]{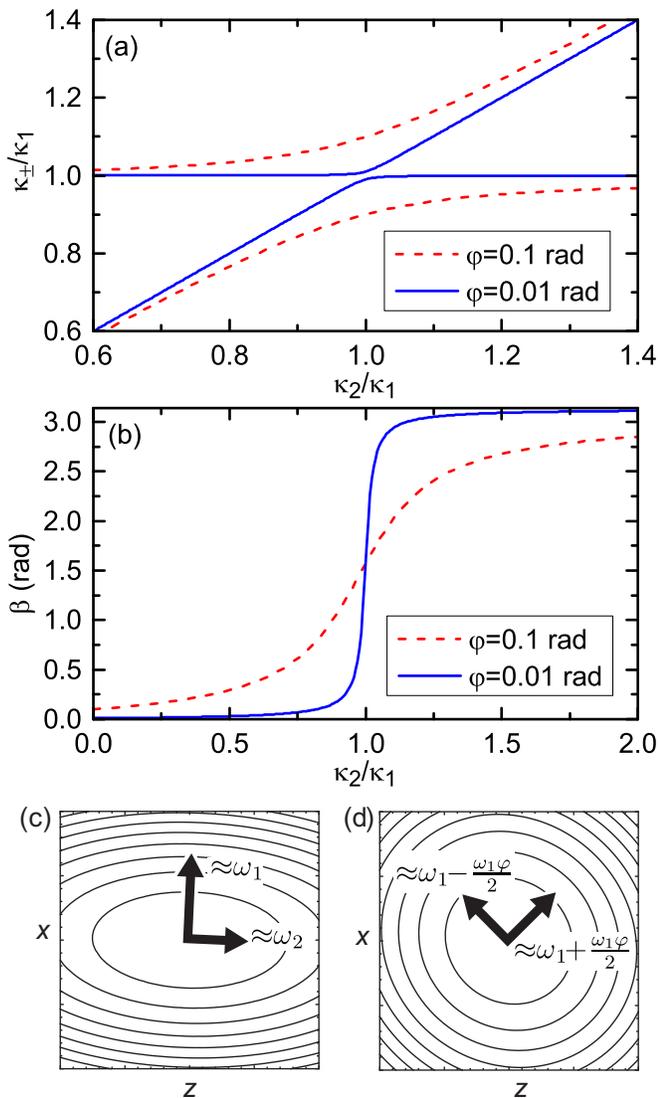}
\caption{
(Color online) (a) Eigenvalues of the spring-constant tensor $\kappa$. The eigenvalues $\kappa_\pm$ of Eq.\ \eqref{kappa-pm} show an avoided crossing as a function of the ratio of lattice spring constants $\kappa_2/\kappa_1$. (b) Rotation of the principal axes. For $\kappa_2/\kappa_1= 0$ and $\kappa_2/\kappa_1\to \infty$, one obtains $\beta= \varphi$ and $\beta\to \pi-\varphi$, respectively. This means that the principal axis $\bm e_+$ is along the tightly confining lattice, as illustrated in (c), where equipotential lines and principal axes of the trap are shown. For $\kappa_2= \kappa_1$, one obtains $\beta= \pi/2$ which means that the principal axes are the angle bisectors of the wave vectors. For $\varphi\ll 1$ and $\kappa_2= \kappa_1$, the principal axes are rotated by $\approx 45^\circ$ relative to the wave vectors, as illustrated in (d). This $45^\circ$ rotation is the central result of our model, because it implies that laser cooling which acts only along one lattice wave vector can address both principal axes.
\label{fig-theory}
}
\end{figure}

Figure \ref{fig-theory} illustrates these results. Part (a) shows the eigenvalues $\kappa_\pm$ at fixed $\varphi$ and $\kappa_1$. They obviously display an avoided crossing as a function of $\kappa_2$. This behavior is expected because the expression $\kappa_+-\kappa_-= \sqrt{[\kappa_2-\kappa_1\cos(2\varphi)]^2+[\kappa_1\sin(2\varphi)]^2}$ is mathematically equivalent to the familiar avoided crossing $\sqrt{(\omega-\omega_\text{res})^2+\omega_R^2}$ with a tunable parameter $\omega$, a resonance position $\omega_\text{res}$, corresponding to $\kappa_1\cos(2\varphi)$, and a coupling term $\omega_R$, corresponding to $\kappa_1\sin(2\varphi)$.

As an aside, the lack of symmetry under swapping the wave vectors in the above result $\kappa_2= \kappa_1\cos(2\varphi)$ for the position of the avoided crossing is a result of keeping $\kappa_1$ fixed, a quantity which is not symmetric under this swap. If we instead fix a symmetric expression, such as $\kappa_1+\kappa_2$, then the minimum splitting will occur for $\kappa_1= \kappa_2$.

Figure \ref{fig-theory}(b) shows that $\beta$ is a strictly increasing function of $\kappa_2/\kappa_1$. In the limit where $\kappa_1$ or $\kappa_2$ vanishes, it is intuitively clear that the principal axes must be along and orthogonal to the remaining lattice wave vector. This is mathematically expressed by the fact that $\beta\to \varphi$ for $\kappa_2/\kappa_1\to0$ so that $\bm e_+\to \bm k_1/k_1$. Similarly, $\beta\to \pi-\varphi$ for $\kappa_2/\kappa_1\to\infty$ so that $\bm e_+\to \bm e_z= \bm k_2/k_2$. This aspect of cross-dimensional mixing in a slightly nonorthogonal 2D lattice has previously been observed in an ultracold atomic gas \cite{Mueller:07}, but not in the context of laser cooling.

The physical situation obtained for $\kappa_1= \kappa_2$ is much more intriguing. It is obviously symmetric under swapping the wave vectors. Hence, it is intuitively clear that the principal axes must be the angle bisectors of the wave vectors. Indeed, assuming $\kappa_1= \kappa_2$ and $\sin\varphi\neq 0$ yields $\beta= \pi/2$. If additionally $\varphi\ll 1$, then the principal axes are rotated by $\approx 45^\circ$ relative to the wave vectors. In this situation, cooling light which acts only along one lattice wave vector can address both principal axes. This was not directly obvious from Eq.\ \eqref{V-k-j}.

For the symmetric configuration, $\kappa_1= \kappa_2$, we obtain trap eigenfrequencies $\omega_\pm= \omega_1\sqrt{1\pm\sin\varphi}$. For $\varphi\ll 1$, this simplifies to
\begin{align}
\label{omega-splitting}
\omega_+-\omega_-
\approx \omega_1\varphi
.\end{align}
The central result of this model is that tuning the ratio of the trap frequencies makes it possible to rotate the principal axes of the combined potential into any desired direction in the lattice plane. Note that the possibility to rotate principal axes with this technique has a straightforward generalization to three dimensions.

Mathematically, the 45$^\circ$ rotation at $\kappa_1= \kappa_2$ exists for any arbitrarily small nonzero value of $\varphi$. In an experiment, imperfections such as technical fluctuations in $\kappa_2/\kappa_1$ will result in a minimum value of $\varphi$ required to make the 2D Raman cooling work stably.

As detailed below, the geometry of our experiment is such that the projection $\bm e_\pm\cdot \bm k_2$ of the wave vector $\bm k_2$ onto the eigenmodes is a crucial parameter for the Raman cooling. To express this parameter in terms of $\varphi$ and $\kappa_2/\kappa_1$, we first use Eqs.\ \eqref{k-1-2-prime} and \eqref{e-pm-prime} to obtain $\bm e_+\cdot \bm k_2= k_2 \sin \frac{\varphi+\beta}{2}$ and $\bm e_-\cdot \bm k_2= k_2 \cos \frac{\varphi+\beta}{2}$ and then use trigonometric identities, which yield
\begin{align}
\label{k-pm}
\left(\frac{\bm e_\pm\cdot\bm k_2}{k_2}\right)^2
= \frac{1\mp\cos(\varphi+\beta)}2
.\end{align}
Inserting Eq.\ \eqref{arg-beta} for $\beta$ yields an expression in terms of $\varphi$ and $\kappa_2/\kappa_1$.

\subsection{Selection Rules in Raman Transitions}

In addition to the two standing waves with wave vectors $\bm k_1$ and $\bm k_2$, we consider an additional traveling light wave with wave vector $\bm k_3$ and angular frequency $\omega_{3,0}$. The combination of the last two light fields is used to drive resolved-sideband Raman transitions between two hyperfine components of the atomic ground state. Let
\begin{align}
\label{Delta-omega}
\Delta\omega
= \omega_{3,0}-\omega_{2,0} +\omega_\text{HF}
,\end{align}
denote the detuning between the angular frequencies of these two light fields relative to the free-space hyperfine splitting $\omega_\text{HF}$. In the presence of the tightly confining 2D lattice, resonant Raman transitions require
\begin{align}
\Delta\omega
= \omega_+\Delta n_+ + \omega_-\Delta n_-
,\end{align}
where $\omega_+$ and $\omega_-$ from Eq.\ \eqref{omega-kappa} are the angular eigenfrequencies of the trap and $n_\pm$ and $n_\pm'= n_\pm+\Delta n_\pm$ denote the vibrational quantum numbers along the $\bm e_\pm$ eigenmodes before and after the Raman transition, respectively. We consider the Lamb-Dicke regime, characterized by the criterion that the Lamb-Dicke parameters $\eta_\pm$ are much below unity \cite{Leibfried:03:RMP}. Hence, the two-photon Rabi frequency for the Raman transition is suppressed by a factor $\eta_+^{|\Delta n_+|}\eta_-^{|\Delta n_-|}$. We assume $\eta_+\approx \eta_-$. Hence, the strongest transition is the carrier transition $\Delta n_+= \Delta n_-= 0$ and the next weaker transitions are the first-order motional sidebands with
\begin{subequations}
\label{Delta-n-plus-minus}
\begin{align}
\Delta n_+= \pm1
&&
\text{ and }
&&
\Delta n_-= 0
\end{align}
or
\begin{align}
\Delta n_+= 0
&&
\text{ and }
&&
\Delta n_-= \pm1
.\end{align}
\end{subequations}
The considerations so far suggest that one could drive the first-order red sideband with $\Delta n_+= -1$ and $\Delta n_-= 0$ to cool the atomic motion along $\bm e_+$ and subsequently the first-order red sideband with $\Delta n_+= 0$ and $\Delta n_-= -1$ to cool the atomic motion along $\bm e_-$. This would allow for 2D Raman cooling.

However, this conclusion does not necessarily hold under all circumstances. To see this, we note that the light with wave vector $\bm k_2$ plays a double role by providing part of the lattice potential and simultaneously contributing to the Raman transition. This has profound consequences. We assume that $\bm k_1$ and $\bm k_3$ are parallel and that the lattice potential with wave vector $\bm k_2$ is repulsive, so that the atom is trapped at a node of this standing wave. With the coordinate origin at this node, the standing-wave electric field amplitude is proportional to $\sin(\bm k_2\cdot\bm x)$. The two-photon Rabi frequency for the Raman transition is, see e.g.\ Ref.\ \cite{Reimann:14},
\begin{align}
\label{omega-two-photon}
\omega_\text{two-ph}
= \omega_0\langle n_+',n_-'|e^{i\bm k_3\cdot \bm x}\sin(\bm k_2\cdot\bm x)|n_+,n_-\rangle
,\end{align}
where $\omega_0$ is the two-photon Rabi frequency in free space, which is determined by the electric dipole matrix element and the intensities and single-photon detuning of the Raman light fields.

To see why the matrix element in Eq.\ \eqref{omega-two-photon} can hamper 2D Raman cooling, we assume $\varphi\ll 1$ and $\omega_1\varphi\ll|\omega_1-\omega_2|$, so that cross-dimensional mixing is negligible. Hence, the normal modes of the atomic motion are approximately along the coordinate axes $x$ and $z$. We denote the corresponding vibrational quantum numbers as $n_1$ and $n_2$. Using $\bm k_2\cdot\bm k_3\approx 0$, we find that $e^{i\bm k_3\cdot \bm x}\sin(\bm k_2\cdot\bm x)$ is an asymmetric function along $\bm e_z= \bm k_2/k_2$. This implies that the symmetry of the vibrational wave function along $z$ must change during the Raman transition. This is expressed by the selection rule \cite{Boozer:08}
\begin{align}
\label{Delta-n2}
\Delta n_2
\text{ must be odd.}
\end{align}
This selection rule obviously removes the carrier. This does not pose a problem for Raman cooling. It might even have the potential to be advantageous regarding the final temperature \cite{Reimann:14}. In addition, this selection rule removes those first-order sidebands of Eq.\ \eqref{Delta-n-plus-minus} which have $\Delta n_1= \pm1$ and $\Delta n_2=0$. Hence, the motion along $\bm e_x\approx \bm k_1/k_1$ cannot be Raman cooled with this scheme and the cooling becomes one dimensional (1D). This can be regarded as complete destructive interference between the components of the matrix element in Eq.\ \eqref{omega-two-photon} which arise from the $e^{ik_2z}$ and $e^{-ik_2z}$ terms, which together form $\sin(k_2z)$.

Reference \cite{Reimann:14} overcame this problem by driving a second-order red sideband with $\Delta n_1=\Delta n_2= -1$. The corresponding two-photon Rabi frequency is suppressed by the square of the Lamb-Dicke parameter, which reduces the cooling rate. Hence, the mean excitation number in the direction which is difficult to cool was experimentally limited to $0.3(2)$ in Ref.\ \cite{Reimann:14}. The goal of our work is to use cross-dimensional mixing with $\varphi\ll 1$ and $\omega_1\approx \omega_2$ to overcome this problem, achieve 2D cooling without resorting to second-order sidebands, and thus reach lower mean excitation number.

Note that the difficulty in achieving 2D Raman cooling is not related to the sign of the detuning of the light with wave vector $\bm k_2$ which plays the double role of trapping and contributing to the Raman transition. If this light were red detuned, then the atoms would be trapped at an antinode and the expression $\sin(\bm k_2\cdot\bm x)$ in Eq.\ \eqref{omega-two-photon} would be replaced by $\cos(\bm k_2\cdot\bm x)$. The selection rule Eq.\ \eqref{Delta-n2} would then read ``$\Delta n_2$ must be even''. Hence, one could easily Raman cool the $x$ direction but not the $z$ direction. So the difficulty in achieving 2D Raman cooling would remain. Moreover, using only a moderate detuning of the cavity-mode lattice is desirable because it is the only way to ensure that the lattice nodes and antinodes coincide with the nodes and antinodes of the resonant cavity mode for a relatively long $z$ distance. For only moderate detuning it is desirable to use a blue lattice because it reduces the spontaneous scattering rate and possible fluctuations of the light shifts due to intra-cavity power fluctuations, simply because the atoms are trapped at a node.

\section{Experimental Setup and Procedure}

\label{sec-setup}

\begin{figure}[!tb]
\includegraphics[width=\columnwidth]{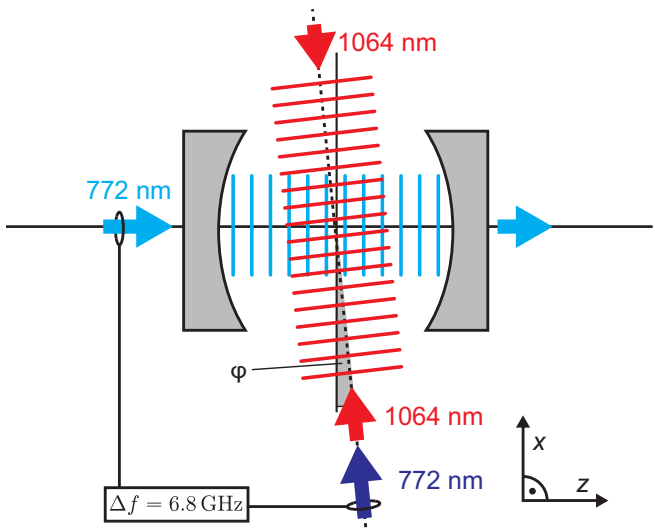}
\caption{
(Color online) A 2D optical lattice is formed inside an optical resonator by intersecting two standing light waves, one (light blue) with a wavelength of 772 nm and a wave vector along the cavity axis ($z$ axis), the other (red) with a wavelength of 1064 nm and forming an angle $\varphi\ll 1$ with the $x$ axis. Aligned parallel with the 1064-nm light, we can apply an additional traveling light wave at 772 nm (dark blue arrow), $\approx6.8$ GHz detuned from the cavity-axis lattice to drive stimulated Raman transitions.
\label{fig-setup}
}
\end{figure}

A scheme of the experimental setup is shown in Fig.\ \ref{fig-setup}. A 2D optical lattice is created by intersecting two standing light waves inside a Fabry-Perot resonator with a finesse of $5.5\times10^4$ and a length of 0.50 mm. The resonator axis is the $z$ axis and serves as a quantization axis for magnetic quantum numbers. A magnetic field of 8.1 $\mu$T applied along the $z$ axis stabilizes the orientation of the atomic spins. Gravity acts along the $y$ axis.

The resonator length is stabilized such that one TEM$_{00}$ resonator mode is near-resonant with the $|5S_{1/2},F{=}2\rangle \leftrightarrow |5P_{3/2},F{=}3\rangle$ cycling transition at 780 nm in $^{87}$Rb. This allows for resonator-enhanced detection of the atomic hyperfine state as described in Ref.\ \cite{Bochmann:10}. In the present paper, the resonator does not serve any purpose beyond enhancing state detection. The polarization modes of the resonator are degenerate, so that any coherent superposition of two basis polarizations can be excited.

The first standing wave (red in Fig.\ \ref{fig-setup}) is created by a free-space laser beam that is retroreflected from a single mirror. This free-space standing wave with wavelength $\lambda_1= 2\pi/k_1= 1064$ nm has a wave vector $\bm k_1$, which lies in the $xz$ plane and forms an angle $\varphi\ll 1$ with the $x$ axis. This light is far red detuned from all resonance lines in $^{87}$Rb, so that it creates an attractive lattice potential with a polarizability of $\alpha_1= 687$ atomic units (a.u.) \cite{Arora:12}, where $1\text{ a.u.}= 1.649\times 10^{-41}$ J(m/V)$^2$. This standing wave has a typical traveling-wave power of 1.3 W and creates a typical trap angular frequency of $\omega_1\approx 2\pi\times 0.5$ MHz along $\bm k_1$. Because of its finite spot size ($1/e^2$ radius of intensity) of $w_1= 16$ $\mu$m, this beam creates an additional confinement along the directions perpendicular to $\bm k_1$ with a trap angular frequency $\sqrt2\omega_1/k_1w_1$ of typically $2\pi\times 8$ kHz. This has little effect along the $z$ axis, where strong confinement is created by the second standing wave, but it creates the dominant confinement along the $y$ axis, supporting the atoms against gravity. The electric field of this standing wave is polarized along the $z$ axis.

The second standing wave (light blue in Fig.\ \ref{fig-setup}) is created by resonantly coupling laser light into one of the TEM$_{00}$ resonator modes. This cavity-mode standing wave with wavelength $\lambda_2= 2\pi/k_2= 772$ nm has its wave vector $\bm k_2$ along the $z$ axis. This light is blue detuned from the D$_1$ and D$_2$ resonance lines at 795 and 780 nm in $^{87}$Rb. It creates a repulsive lattice potential with a polarizability of $\alpha_2= -1.2\times10^4$ a.u.\ \cite{Neuzner:15} and a trap angular frequency $\omega_2$ along $z$, with a value of typically $\omega_2\approx \omega_1$. The power transmitted through the cavity is typically $P_z= 6.5$ $\mu$W. At a power transmission coefficient of $1.0\times10^{-4}$ for the output mirror of the cavity, this corresponds to an intra-cavity traveling-wave power of 65 mW. This optical lattice creates the dominant confinement along $z$. Its spot size of $w_2= 30$ $\mu$m creates a repulsive potential along the directions perpendicular to the $z$ axis. But for a sufficiently cold atom this has negligible effect, because the atom is trapped near a node.

We apply an additional traveling-wave light field at $\lambda_3=772$ nm (dark blue in Fig.\ \ref{fig-setup}). This light field has a wave vector $\bm k_3$ aligned parallel to the 1064-nm light and an electric field polarized along the $y$ axis. It is overlapped with and separated from the 1064-nm beam using dichroic mirrors (not shown in Fig.\ \ref{fig-setup}). It has a frequency, which is approximately $\omega_\text{HF}$ red detuned from the 772-nm lattice light, where $\omega_\text{HF}/2\pi\approx 6.8$ GHz \cite{bize:99} is the ground-state hyperfine splitting in $^{87}$Rb. The combination of these two 772-nm light fields drives stimulated Raman transitions between the $|F{=}1,m_F{=}0\rangle$ and $|F{=}2,m_F{=}0\rangle$ components of the $5S_{1/2}$ ground state. The traveling-wave 772 nm light has a spot size of 34 $\mu$m and a typical power $P_\text{trvl}$ of 3 mW for Raman cooling and between 1 and 6 $\mu$W for spectroscopy. The free-space two-photon Rabi frequency $\omega_0$ would be maximized, if the 772-nm lattice light were linearly polarized along the $x$ axis. This would yield $\omega_0/2\pi= 15$ kHz at $P_z= 6.5$ $\mu$W and $P_\text{trvl}=1$ $\mu$W. In our experiment, the 772-nm lattice light is elliptically polarized, which reduces $\omega_0$.

Finally, we can apply another retroreflected light field (not shown in Fig.\ \ref{fig-setup}) with a wavelength near the D$_2$ line. The wave vector of this light lies in the $xy$ plane and forms an angle of $\approx45^\circ$ with the $x$ axis. This light field has a lin $\perp$ lin polarization configuration and is used for polarization-gradient cooling \cite{dalibard:89}. It is 32 MHz red detuned from the free-space $|5S_{1/2},F{=}2\rangle \leftrightarrow |5P_{3/2},F{=}3\rangle$ transition. For a ground-state atom, the 1064-nm optical lattice has a depth of $V_{1,0}\approx -2\pi\hbar\times 31$ MHz. As a cold atom is localized near an antinode of this light, this level shift is seen by a ground-state atom. According to Ref.\ \cite{Neuzner:15}, the $5P_{3/2}$ state experiences a shift with opposite sign and similar order of magnitude. Both shifts complicate the situation. Nevertheless, the polarization gradient cooling is seen to work well. Whenever we apply this light field, we add repumping light which is 52 MHz blue detuned from the free-space $|5S_{1/2},F{=}1\rangle \leftrightarrow |5P_{3/2},F{=}2\rangle$ transition. The exact value of the detuning is of little relevance as long as it is large enough that different light shifts at different positions in the trap do not drastically alter the repumping rate.

We use an enhanced-mode charge-coupled-device camera to record images of the light, which the atoms emit during polariza\-tion-gradient cooling. The imaging system has a spatial resolution of 1.3 $\mu$m full width at half maximum (FWHM). Using the good signal-to-noise ratio obtained when recording a large number of photons per atom, we determine the relative position of two atoms with a precision of 0.1 $\mu$m FWHM \cite{Neuzner:16:NatPhoton}, which clearly yields single-site resolution. The camera images show the $xz$ plane and yield $\varphi= 27.0(3)\text{ mrad}= 1.55(2)^\circ$ \cite{Neuzner:16:NatPhoton}. Only statistical uncertainties are quoted throughout this work. For the value of $\varphi$ quoted here systematic deviations are most likely much larger than the statistical uncertainty. Such systematic deviations can be caused e.g.\ by geometric image deformations due to optical aberrations combined with possible off-axis imaging.

The experiment begins with the deterministic loading of a single $^{87}$Rb atom into the 2D lattice potential, similar to Ref.\ \cite{Neuzner:16:NatPhoton}. To this end, a small ensemble of atoms is loaded into a magneto-optical trap, transferred into an optical dipole trap, transported into the resonator volume, and then transferred into the 2D optical lattice. After this transfer, the average atom number is roughly 2. These atoms are illuminated with polarization-gradient cooling light and repumping light. The resulting fluorescence light is imaged with the camera. Based on the camera image, a computer program decides, which atom to keep. If there are other atoms, they will all be removed from the lattice, one after the other, using photon recoils from a tightly focused push-out beam. After this procedure exactly one atom is left in the 2D lattice.

The decision, which atom to keep, is made based on the criterion that the atom must lie in a given interval along $z$ with a length of roughly 5 $\mu$m. The low atomic temperature combined with the small spot size of the 1064-nm light creates a narrow atomic position distribution along $z$ before the push-out. Hence, loading events that yield no eligible atom are rare. In those events, the loading procedure is repeated immediately.

After the image-and-push-out procedure, the remaining single atom is positioned along $x$ with a precision of 0.5 $\mu$m by tilting a glass plate which is located in the 1064-nm beam between the atom and the retro-reflecting mirror. The desired $x$ position is chosen at the center of the $z$ lattice beam. After loading and positioning, optical pumping is used to prepare the atom in the $|F{=}1,m_F{=}0\rangle$ substate of the $5S_{1/2}$ ground state.

Once a single atom has been loaded, positioned and optically pumped, the actual experimental sequence with Raman spectroscopy, or Raman cooling etc.\ is repeated every 2 ms. Each repetition is followed by polarization-gradient cooling with subsequent optical pumping. Part of the resulting fluorescence light is collected with the camera. Here, the exposure time for a single camera image is set to 750 ms. Whenever a new camera image is available, we use it to determine the atomic position. Hopping of an atom to a different lattice site is a rare event. The typical time between two hopping events along $x$ is roughly 30 s. If a hopping event along $x$ is detected, the data recorded during the last 750 ms will be ignored and the atom will be repositioned along $x$. If the atom has left the desired $z$ interval by hopping along $z$ or if the atom has left the trap, the data recorded during the last 750 ms will be ignored and a new atom will be loaded.

\section{Raman Spectroscopy}

\label{sec-spectroscopy}

Before turning to Raman cooling, we study Raman spectroscopy to explore which sidebands are available for Raman cooling. The combination of the standing wave 772 nm light with wave vector $\bm k_2$ and the traveling wave 772 nm light with wave vector $\bm k_3$ is used to drive Raman transitions on the $\Delta m_F{=} 0$ clock transition between the $|F{=}1,m_F{=}0\rangle$ and $|F{=}2,m_F{=}0\rangle$ components of the $5S_{1/2}$ ground state. Because of selection rules, this transition would be dipole forbidden, if both light fields were linearly polarized along the same direction. We circumvent this by choosing a linear polarization along the $y$ axis for the traveling wave 772 nm light together with an elliptic polarization of the standing wave 772 nm light.

To induce Raman transitions, we leave the standing wave 772 nm light on continuously. It creates the optical lattice along $z$ and contributes to the Raman transition. The traveling-wave 772-nm light is switched on with a rectangular envelope for 0.3 ms to drive the Raman transition. After this Raman pulse, cavity-enhanced hyperfine state detection \cite{Bochmann:10} is used to determine whether a population transfer into the $F{=}2$ state has occurred. The difference between the trap frequencies $\omega_1/2\pi\approx 530$ kHz and $\omega_2/2\pi\approx 430$ kHz is large enough that cross-dimensional mixing is negligible.

\begin{figure}[!tb]
\includegraphics[width=\columnwidth]{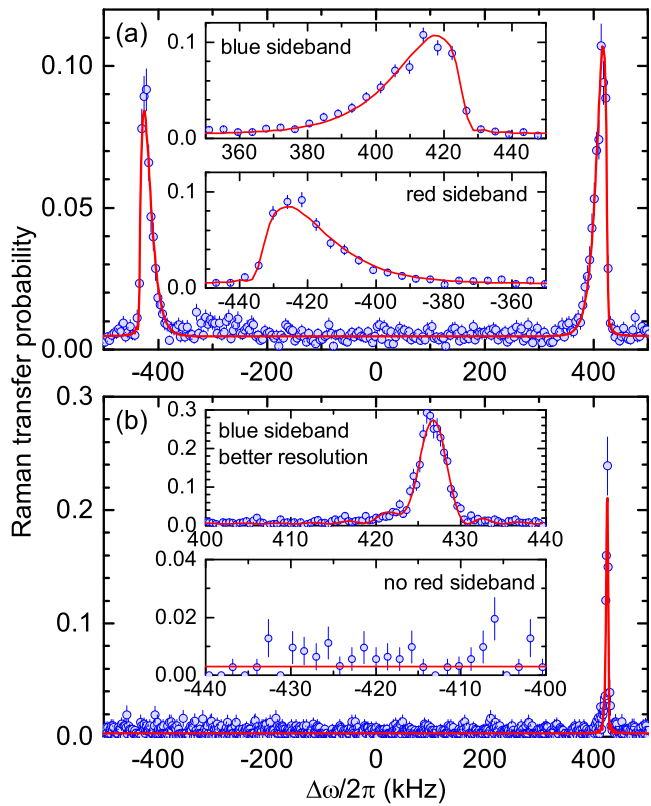}
\caption{
(Color online) Spectroscopy of a carrierless Raman transition. (a) The mechanical selection rule Eq.\ \eqref{Delta-n2} drastically suppresses the carrier at $\Delta\omega\approx 0$. Raman transfer of the atomic population is observed for the sidebands at $\approx \pm430$ kHz, corresponding to vibrational (de-)excitation along the cavity axis $z$. From the areas under the two sidebands, we extract a mean excitation number of $\overline{n_2}= 3.3(2)$. Insets: Close-ups of the two sidebands show asymmetric broadening. (b) Similar spectrum after 1D Raman cooling. The red sideband at $\approx -430$ kHz has vanished, demonstrating successful 1D ground state cooling along $z$. The insets, with different scales on the horizontal and vertical axes compared to (a), show no discernable red sideband and a much-narrowed blue sideband that has only a small residual asymmetry. The data yield $\overline{n_2}< 0.05$. The lines serve to guide the eye.
\label{fig-spectra}
}
\end{figure}

A scan of the Raman detuning $\Delta\omega$ yields the Raman spectrum shown in Fig.\ \ref{fig-spectra}(a). The carrier is predicted to be absent because of the selection rule Eq.\ \eqref{Delta-n2}. Indeed, the data show no evidence of a carrier resonance. However, the first-order red and blue sidebands with $\Delta n_1=0$ and $\Delta n_2=\pm1$ are clearly visible.

Assuming thermal equilibrium, the standard method \cite{Bergquist:87, Leibfried:03:RMP} for estimating the atomic temperature $T$ in Raman spectroscopy is based on a measurement of the areas $A_\text{red}$ and $A_\text{blue}$ under the red and blue sideband, respectively, which yields $A_\text{red}/A_\text{blue}= e^{-\hbar\omega_2/k_BT}$ with the Boltzmann constant $k_B$, see Eqs.\ \eqref{app-q} and \eqref{app-Ared-Ablue} from appendix \ref{sec-app-areas}. After subtracting a background, which we determine from the data far away from the sidebands, we extract $A_\text{red}$ and $A_\text{blue}$ from Fig.\ \ref{fig-spectra}(a). This translates into a mean excitation number of $\overline{n_2}= (A_\text{blue}/A_\text{red}-1)^{-1}= 3.3(2)$ and a temperature of $T= 78(4)$ $\mu$K along the $z$ axis.

Using Eq.\ \eqref{app-omega-0} from appendix \ref{sec-app-areas}, the data in Fig.\ \ref{fig-spectra}(a) also yield $\omega_0/2\pi= 5(1)$ kHz. The data were taken at $P_\text{trvl}= 1$ $\mu$W. The obtained value for $\omega_0$ is a factor $\approx 3$ smaller than the estimate in Sec.\ \ref{sec-setup} which is oversimplified by assuming that the 772-nm lattice light would be linearly polarized along $x$. We believe that most of this factor is explained by the elliptic polarization of the 772-nm lattice light, which is expected to reduce $\omega_0$.

Both sidebands show asymmetric broadening with the shallow sideband edge facing the carrier, as previously observed e.g.\ in Refs.\ \cite{Blatt:09, McDonald:15}. This is caused by the nonzero atomic temperature in all three dimensions which makes the atom sample anharmonic terms of the trapping potential. Hence, the linewidth can be used to estimate the temperature, somewhat similar to e.g.\ Ref.\ \cite{Blatt:09}. We do not detail this here. However, we note that each sideband in Fig.\ \ref{fig-spectra}(a) has a width of $\Delta\omega_\text{FWHM}/2\pi\approx 20$ kHz.

The asymmetric broadening caused by different polarizabilities \cite{McDonald:15} is negligible here because the polarizabilities of the $F{=}1$ and $F{=}2$ ground states are very similar. This is immediately evident from the data because if this effect would dominate, the shallow sideband edges would appear on the same (e.g.\ red-detuned) side of each sideband. However, these different polarizabilities cause a common shift of all lines by a few kilohertz, which is barely visible in some of our data.

\section{1D Raman Cooling}

Figure \ref{fig-spectra}(b) shows a spectrum taken with the same trap settings as in part (a) but after application of 1D Raman cooling. The cooling is achieved by alternating Raman transfer on the red sideband at $\approx -430$ kHz with optical pumping back into the original internal state. As the experiment is deeply in the Lamb-Dicke regime, the optical repumping leaves the vibrational state unchanged with high probability. Hence, the net effect of one cooling cycle is to remove exactly one quantum of vibrational energy along $z$. For experimental convenience, both Raman light fields are on continuously and only the repumping light is pulsed. We use a combination of hyperfine and Zeeman repumping light. One repumping interval lasts 9.5 $\mu$s and is repeated every 15 $\mu$s. Between the application of repumping light, the Raman transition is driven coherently for $t=5.5$ $\mu$s. Hence, pure interaction-time broadening at $T=0$ according to Eq.\ \eqref{app-P-vs-Delta-approx} predicts a FWHM of $2\pi\times0.886/t= 2\pi\times 160$ kHz addressed during a single cooling pulse. In the spectroscopy in Fig.\ \ref{fig-spectra}(a), which uses longer Raman pulses, the red sideband has a width which is considerably smaller than 160 kHz. Hence all population participating in the red sideband is expected to be addressed simultaneously by the cooling light.

Figure \ref{fig-spectra}(b) was recorded after 21 cycles of 1D Raman cooling. The lower inset shows the part of this data set in which the red sideband would be expected to occur near $-430$ kHz. There is no discernable red sideband, proving successful cooling into the vibrational ground state along $z$. Data analysis along the same lines as for part (a) yields $\omega_0/2\pi= 7(1)$ kHz and shows that $\overline{n_2}$ and $T$ are compatible with zero with 68\% confidence intervals of roughly $\overline{n_2}< 0.05$ and $T<7$ $\mu$K.

The upper inset shows another data set recorded with the same parameters but smaller step size in $\Delta\omega$, showing only the blue sideband. From these data we extract a width of the blue sideband of $\Delta\omega_\text{FWHM}\approx 2\pi\times 3.7$ kHz, which is comparable to the expected interaction-time broadening of $2\pi\times 3.0$ kHz for a pulse length of 0.3 ms used during spectroscopy. This shows that after cooling, interaction-time broadening dominates the spectra over other broadening mechanisms. Note that we deliberately use relatively short cooling pulses to achieve cooling for a broad frequency range, whereas we use much longer Raman pulses for the subsequent spectroscopy to resolve narrow features.

\section{Cross-Dimensional Mixing}

We record a series of spectra like the one shown in Fig.\ \ref{fig-spectra}(b) for different values of the trap depth along the cavity axis $z$, keeping the other trap parameters constant. The timing and power of the applied Raman cooling light is also the same as in Fig.\ \ref{fig-spectra}(b).

\begin{figure}[!t]
\includegraphics[width=\columnwidth]{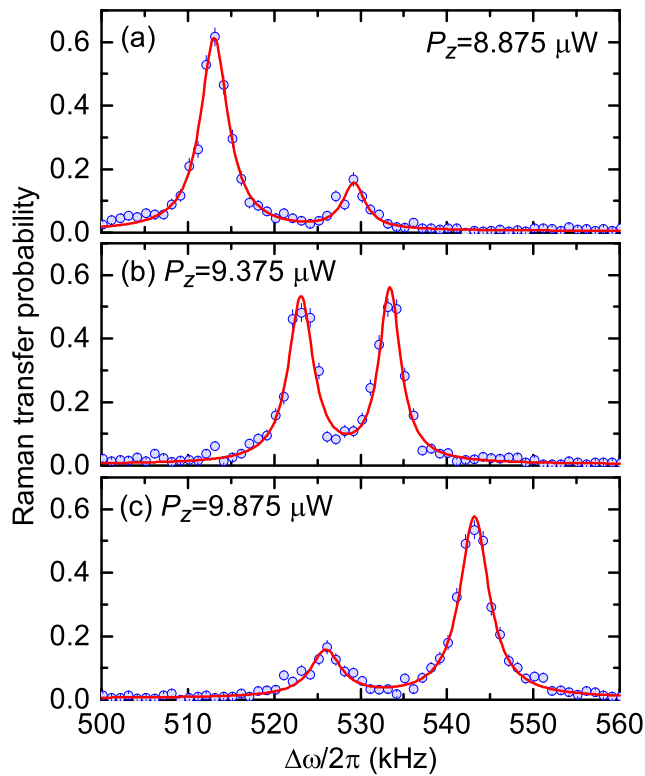}
\caption{
(Color online) Examples of spectra recorded for near-degenerate trap frequencies. The lines show a fit of the sum of two Lorentzians. As a result of cross-dimensional mixing, two blue sidebands are visible in each of these three spectra. In part (b) $\omega_1\approx\omega_2$, whereas parts (a) and (c) are for slightly smaller and larger lattice depth of the cavity-mode standing wave, respectively. This affects the resonance frequencies as well as the areas under the sidebands.
\label{fig-crossing-spectra}
}
\end{figure}

The blue-sideband parts of three example spectra are shown in Fig.\ \ref{fig-crossing-spectra}. In each part of the figure, the line shows a fit of the sum of two Lorentzians plus an offset. This fit yields the center frequency, area, and width of each sideband. For $\omega_1\approx \omega_2$ we expect cross-dimensional mixing and, indeed, we observe two blue sidebands in this regime, as clearly seen in the figure.

\begin{figure}[!t]
\includegraphics[width=\columnwidth]{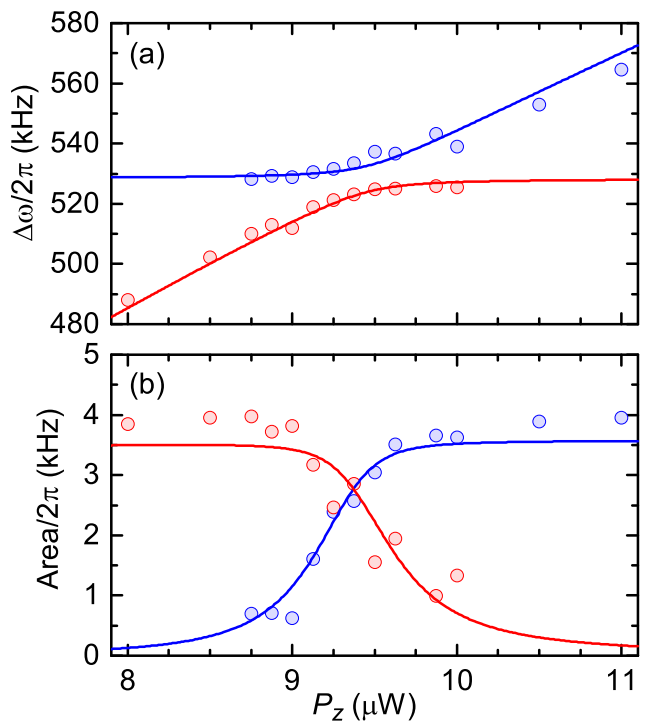}
\caption{
(Color online) Observation of cross-dimensional mixing. (a) Eigenfrequencies of the two blue sidebands extracted from fitting to a set of spectra similar to Fig.\ \ref{fig-crossing-spectra}. The avoided crossing expected from Fig.\ \ref{fig-theory}(a) is clearly visible. (b) Corresponding areas under the two blue sidebands. According to Eqs.\ \eqref{A-10-pm-cdot}--\eqref{chi}, the areas are essentially given by the projection of the eigenmodes onto the $z$ axis. Hence, the change of the areas demonstrates the rotation of the eigenmodes. The vanishing area of the blue sideband at $\approx 530$ kHz when moving far away from the crossing point is a manifestation of the mechanical selection rule Eq.\ \eqref{Delta-n2}.
\label{fig-crossing}
}
\end{figure}

The best-fit values for the center frequencies of both blue sidebands and for the area under both blue sidebands are shown in Fig.\ \ref{fig-crossing}(a) and (b), respectively, as a function of the 772-nm light power $P_z$ transmitted through the cavity, which is proportional to the cavity-field dipole potential depth $V_{2,0}$. The avoided crossing predicted in Fig.\ \ref{fig-theory}(a) is clearly visible in Fig.\ \ref{fig-crossing}(a), except now with different scalings on both axes. For the two left-most and the two right-most data points, the signal of one of the two blue sidebands was so close to the experimental noise that we were unable to reliably extract center frequency and area.

The lines in Fig.\ \ref{fig-crossing}(a) show the results of a fit of $\omega_+/2\pi$ and $\omega_-/2\pi$ to the data, using Eqs.\ \eqref{kappa-pm}, \eqref{omega-kappa}, and
\begin{align}
\label{P0}
\omega_2(P_z)
= \omega_1\sqrt{\frac{P_z}{P_0}}
,\end{align}
where $P_0$ is the value of $P_z$ at which the avoided crossing occurs. The fit is performed for the red and blue data points simultaneously. The best-fit values are $\varphi= 16(3)$ mrad, $\omega_1/2\pi= 528(1)$ kHz, and $P_0= 9.5(1)$ $\mu$W. According to Eq.\ \eqref{omega-splitting}, the minimum splitting is $\omega_1\varphi= 2\pi\times 9(2)$ kHz.

Modeling the areas under the blue sidebands in the zero-temperature limit, as detailed in appendix \ref{sec-app-areas}, yields
\begin{align}
\label{A-10-pm-cdot}
A_{1,0,\pm}
= \frac\pi{2t} \vartheta_{1,0,\pm}^2 \; \chi(\vartheta_{1,0,\pm})
\end{align}
with the pulse area
\begin{align}
\label{theta-10-pm-cdot}
\vartheta_{1,0,\pm}
= t \omega_\text{two-ph}
= \sqrt{\frac{\omega_1}{\omega_\pm} \frac{P_z}{P_1} \left(\frac{\bm e_\pm\cdot\bm k_2}{k_2}\right)^2 }
,\end{align}
where $P_1$ is a parameter which is independent of $P_z$. In addition, we abbreviated
\begin{align}
\label{chi}
\chi(\vartheta)
= \frac\pi2 J_1(\vartheta)H_0(\vartheta) +J_0(\vartheta)\left(1-\frac\pi2 H_1(\vartheta)\right)
\end{align}
with the Struve functions $H_n$ and with the Bessel functions of the first kind $J_n$. The global maximum of $\chi(\vartheta)$ is $\chi(0)= 1$. Hence, in the limit of small pulse area we obtain the simple relation $A_{1,0,\pm}= (\pi/2t) \vartheta_{1,0,\pm}^2$ whereas for large pulse area the value of $A_{1,0,\pm}$ is suppressed by an additional factor $\chi(\vartheta_{1,0,\pm})< 1$.

In the absence of cross-dimensional mixing, the eigenmode which is essentially along $\bm e_x$ will have negligible projection onto $\bm k_2/k_2= \bm e_z$. In this regime, the pulse area $\vartheta_{1,0,\pm}$ and the area under the sideband $A_{1,0,\pm}$ are both negligible so that this eigenmode cannot be addressed in Raman cooling.

The lines in Fig.\ \ref{fig-crossing}(b) show the results of a simultaneous fit of this model to all the data in Fig.\ \ref{fig-crossing}(b) using Eqs.\ \eqref{kappa-pm}, \eqref{omega-kappa}, \eqref{arg-beta}, \eqref{k-pm}, and \eqref{P0}--\eqref{chi}. The duration $t=0.3$ ms of the spectroscopy pulse is known prior to fitting. Hence, the model has three free fit parameters. The best-fit values are $\varphi= 22(3)$ mrad, $P_0= 9.4(1)$ $\mu$W, and $P_1= 0.9(1)$ $\mu$W. The best-fit values for $\varphi$ and $P_0$ extracted from Figs.\ \ref{fig-crossing}(a) and (b) agree fairly well with each other.

As mentioned above, the value of $\varphi= 27$ mrad extracted from the camera images described in Sec.\ \ref{sec-setup} most likely suffers from systematic deviations. In addition, the exact value of $\varphi$ is of little relevance for Raman cooling. The only aspect that matters is that $\varphi$ is large enough that technical fluctuations e.g.\ in $\omega_1/\omega_2$ do not compromise the stability at which the 2D Raman cooling works.

When $\omega_1$ and $\omega_2$ differ strongly, one eigenmode has a frequency of $\approx \omega_1$ and is oriented essentially along the $x$ axis. The area under the corresponding sideband vanishes, which is a manifestation of the mechanical selection rule Eq.\ \eqref{Delta-n2}. This is the reason why the Raman cooling is 1D in this regime. For $\omega_1\approx\omega_2$, however, the cross-dimensional mixing makes the areas under the two blue sidebands similar, as clearly seen in Figs.\ \ref{fig-crossing-spectra}(b) and \ref{fig-crossing}(b). This makes both eigenmodes accessible for Raman transitions and 2D Raman cooling becomes possible. This is the central finding of this work.

\begin{figure}[!t]
\includegraphics[width=\columnwidth]{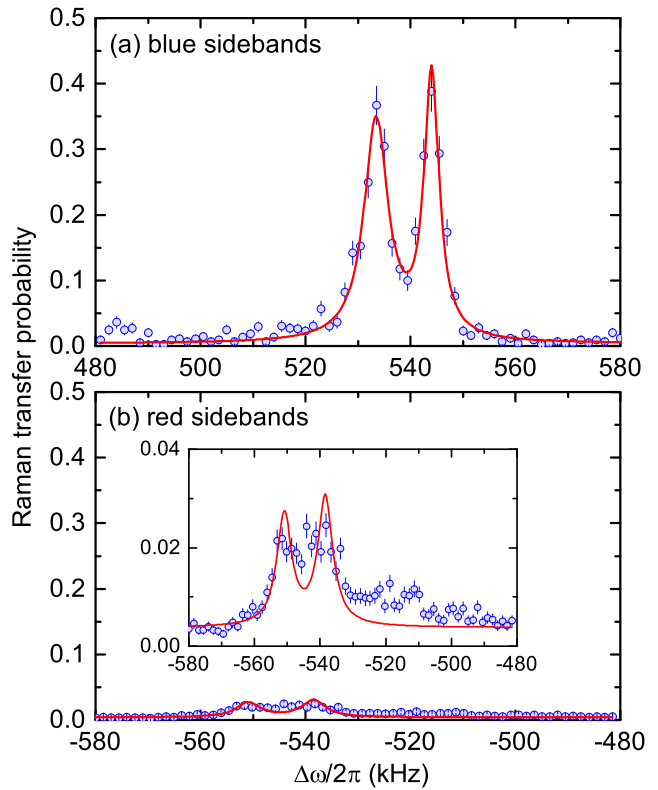}
\caption{
(Color online) Determination of the mean excitation number from a spectrum after 2D Raman cooling for $\omega_1\approx\omega_2$. The strong suppression of both red sidebands compared to both blue sidebands clearly demonstrates low temperature in 2D. Each line shows a fit of the sum of two Lorentzians to one data set. The inset shows a close-up. Depending on the data analysis method (see text), we extract mean excitation numbers of $\overline{n_+}= \overline{n_-}= 0.09(1)$ or $\overline{n_+}= \overline{n_-}= 0.14(1)$.
\label{fig-red-sidbands-2D}
}
\end{figure}

\section{2D Raman Cooling}

Figure \ref{fig-crossing-spectra}(b) clearly shows that both blue sidebands are narrow, which shows that the Raman cooling used here does work as a 2D process. To quantify the low temperature in 2D, we record a spectrum like in Fig.\ \ref{fig-crossing-spectra}(b), but this time including the red sidebands.

This spectrum is shown in Fig.\ \ref{fig-red-sidbands-2D}. The cooling scheme used here does not address each of the two red sidebands individually with resonant light but rather uses only one light field at a detuning of $\Delta\omega/2\pi= -545$ kHz, approximately halfway between the two red sidebands. With an estimated interaction-time broadening of $2\pi\times 160$ kHz, this is expected to address all population on both red sidebands. The line in (a) shows a fit of the sum of two independent Lorentzians plus an offset. The data in part (b) show that both red sidebands are strongly suppressed, proving low temperature in 2D. The line shows a fit of two Lorentzians with a fixed vertical offset taken from (a) and fixed widths of $2\pi\times 5.5$ kHz (FWHM) taken from one of the peaks in (a). The total area $A_\text{red}$ under both peaks of the fit curve in (b) combined with the analogous value $A_\text{blue}$ from (a) together with the assumption of thermal equilibrium in the $xz$ plane yields $T= 10.4(5)$ $\mu$K and $\overline{n_+}= \overline{n_-}= 0.09(1)$.

While the fit in (b) clearly captures the major feature of the red sidebands, there seems to be an additional signal on the shallow sideband edge facing the carrier, which this simple fit does not attempt to model. The physical origin of this signal is unclear. As an alternative method for determining $A_\text{red}$, we therefore directly calculate the area under the raw data points, subtracting the known offset from part (a). Again using the value $A_\text{blue}$ from the fit curve in (a) and assuming thermal equilibrium in the $xz$ plane, this yields $T= 12.1(4)$ $\mu$K and $\overline{n_+}= \overline{n_-}= 0.14(1)$.

\begin{figure}[!t]
\includegraphics[width=\columnwidth]{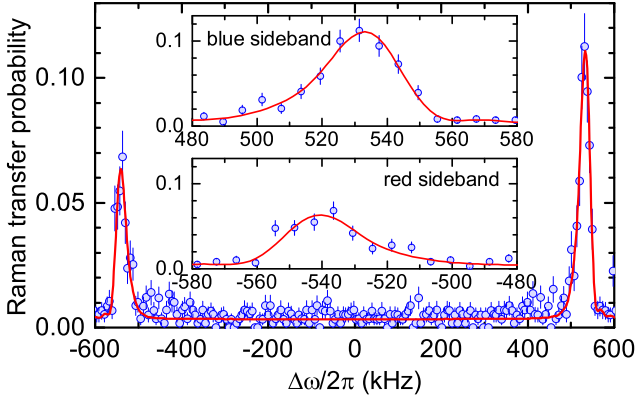}
\caption{
(Color online) Raman spectrum for $\omega_1\approx \omega_2$ without Raman cooling. The sidebands are thermally broadened so strongly that the splittings of the red and blue sidebands seen in Fig.\ \ref{fig-red-sidbands-2D} are not visible here. From the areas under the sidebands, we extract $\overline{n_+}= \overline{n_-}= 1.3(3)$. Comparison with Fig.\ \ref{fig-red-sidbands-2D} shows that the low 2D temperature observed in Fig.\ \ref{fig-red-sidbands-2D} is the result of 2D Raman cooling. The lines serve to guide the eye.
\label{fig-uncooled}
}
\end{figure}

For reference, Fig.\ \ref{fig-uncooled} shows a Raman spectrum taken for degenerate trap frequencies and without Raman cooling. This spectrum clearly shows a strong red sideband, with $A_\text{red}/A_\text{blue}$ yielding $T= 45(8)$ $\mu$K and $\overline{n_+}= \overline{n_-}= 1.3(3)$. In addition, the red and blue sideband in this spectrum have a FWHM of $\approx 2\pi\times 27$ kHz and a clearly asymmetric line shape. This asymmetric broadening is so large that the $\approx 2\pi\times 10$ kHz splittings between the two blue sidebands and between the two red sidebands are not resolved. Both the ratio between sideband areas and the width of the asymmetric broadening show that the temperature before Raman cooling is high. Combining this with the low mean excitation number reached in Fig.\ \ref{fig-red-sidbands-2D}, we conclude that the Raman cooling actually works in 2D.

\section{Conclusions}

We studied a system in which limited optical access makes it difficult to implement Raman cooling in more than 1D. We use a 2D optical lattice in which the lattice beams are slightly nonorthogonal and carefully tune the trapping frequencies created by the two lattice beams to be identical. This spatially rotates the principal axes of the atomic motion in the lattice by 45$^\circ$, thus making it possible to realize 2D Raman cooling, in which we achieve large ground-state occupation in 2D.

\acknowledgments

This work was supported by the European Union via Seventh Framework Programme Collaborative Project SIQS, by the Bundesministerium f\"ur Bildung und Forschung via IKT 2020 (Q.com-Q), and by Deutsche Forschungsgemeinschaft via NIM.

\appendix

\section{Areas under Sidebands}

\label{sec-app-areas}

A model for the areas in Fig.\ \ref{fig-crossing}(b) is obtained easily from Eq.\ \eqref{omega-two-photon}. Deeply in the Lamb-Dicke regime, we neglect terms of order $O(x^2)$ and obtain
\begin{align}
\frac{\omega_\text{two-ph}}{\omega_0}
= \langle n_+',n_-'|\bm k_2\cdot\bm x|n_+,n_-\rangle
.\end{align}
We expand the vectors $\bm k_2$ and $\bm x$ in the orthonormal basis $(\bm e_+,\bm e_-,\bm e_y)$ given by Eq.\ \eqref{e-pm-prime}. This yields coordinate tuples $(k_+,k_-,k_y)$ and $(x_+,x_-,y)$. Note that this yields $k_\pm= \bm e_\pm\cdot\bm k_2$ and $\bm k_2\cdot\bm x= k_+x_++k_-x_-+k_yy$. For the geometry of our experiment $k_y=0$. Hence
\begin{multline}
\label{app-omega-two-ph}
\frac{\omega_\text{two-ph}}{\omega_0}
= \langle n_+'|k_+x_+|n_+\rangle\delta_{n_-',n_-}
\\
+\langle n_-'|k_-x_-|n_-\rangle\delta_{n_+',n_+}
,\end{multline}
We are interested in the first-order blue sideband, i.e.\ $n_+'= 1+n_+$, and obtain
\begin{align}
\label{app-matrix-el-plus}
\langle 1+n_+|k_+x_+|n_+\rangle
= \eta_+ \sqrt{1+n_+}
\end{align}
with the Lamb-Dicke parameter $\eta_+= k_+\sqrt{\hbar/2m\omega_+}= (k_+/k_2) \sqrt{\omega_\text{rec,2}/\omega_+}$, where $\omega_\text{rec,2}= \hbar k_2^2/2m= 2\pi\times 3.85$ kHz is the recoil angular frequency of the 772-nm light. Using $k_2= 2\pi/772$ nm and $\omega_z\approx 2\pi\times 500$ kHz and $\omega_x\approx\omega_z$, which implies $\omega_+\approx\omega_z$ we obtain $\eta_+\approx 0.09 (k_+/k_2)$. According to Eq.\ \eqref{k-pm} $(k_+/k_2)^2$ can vary between 0 and 1. There are analogous expressions for $\langle 1+n_-|k_-x_-|n_-\rangle$ and $\eta_-$.

A Raman pulse with a duration of $t=0.3$ ms drives Rabi oscillations between the $|F{=}1,m_F{=}0,n\rangle$ and $|F'{=}2,m_F'{=}0,n'\rangle$ components of the $5S_{1/2}$ ground state, where $n$ and $n'$ denote the values of the relevant vibrational quantum number before and after the light pulse, respectively. Hence, the Raman transfer probability is
\begin{align}
\label{app-P-exact}
P_{n',n}(\Delta_R)
= \frac{\omega_{n',n}^2}{\omega_{n',n}^2+\Delta_R^2} \sin^2\left(\frac t2 \sqrt{\omega_{n',n}^2+\Delta_R^2}\right)
,\end{align}
where $\omega_{n',n}$ is the two-photon Rabi frequency and $\Delta_R$ the two-photon detuning from the $n\to n'$ Raman resonance.

\begin{figure}[!t]
\includegraphics[width=\columnwidth]{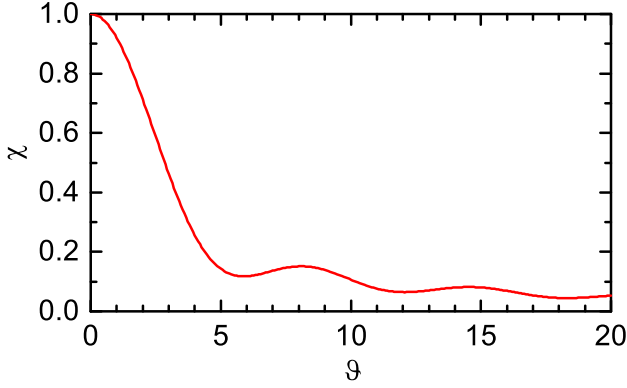}
\caption{
(Color online) The function $\chi(\vartheta)$ defined in Eq.\ \eqref{chi}. For small pulse area $\vartheta\ll1$ the function yields 1 and the area under the sideband is described by the simple expression $A_{n',n}= \pi \omega_{n',n}^2 t/2$. If $\vartheta$ is not small, then the area $A_{n',n}$ will be a factor $\chi(\vartheta)$ smaller than this simple expression.
\label{fig-chi}
}
\end{figure}

We define the pulse area $\vartheta$ of the Raman pulse used for spectroscopy
\begin{align}
\vartheta
= \omega_{n',n} t
.\end{align}
We use this to write the area under the curve $P_{n',n}(\Delta_R)$ as
\begin{align}
\label{app-A-n'n}
A_{n',n}
= \int_{-\infty}^\infty P_{n',n} d\Delta_R
= \frac\pi{2t} \vartheta^2 \; \chi(\vartheta)
,\end{align}
where we abbreviated
\begin{align}
\chi(\vartheta)
= \frac4{\pi\vartheta} \int_1^\infty du \frac{\sin^2\frac{\vartheta u}2}{u\sqrt{u^2-1}}
\end{align}
and where we substituted $\Delta_R$ by $u= t\sqrt{\omega_{n',n}^2+\Delta_R^2}/\vartheta$. This integral can be solved analytically, yielding Eq.\ \eqref{chi}. $\chi(\vartheta)$ is displayed in Fig.\ \ref{fig-chi}. For small $\vartheta$ one obtains $\chi(\vartheta)= 1-\frac1{12}\vartheta^2+O(\vartheta^4)$. According to Eq.\ \eqref{app-A-n'n}, the area $A_{1,0}$ under the blue sideband corresponding to $\omega_\pm$ in the zero-temperature limit is given by Eq.\ \eqref{A-10-pm-cdot}.

We use Eqs.\ \eqref{app-omega-two-ph} and \eqref{app-matrix-el-plus} with $\omega_{n',n}= \omega_\text{two-ph}$ to obtain the pulse area
\begin{align}
\label{app-theta-10-pm-eta}
\vartheta_{1,0,\pm}
= \omega_0 t \eta_\pm
= \omega_0 t \frac{k_\pm}{k_2} \sqrt{\frac{\omega_\text{rec,2}}{\omega_\pm}}
.\end{align}
In the measurements shown in Fig.\ \ref{fig-crossing}(b), we use the same power $P_z$ of the 772-nm cavity light field to create the lattice and to probe the Raman transfer probability. Hence $\omega_0\propto \sqrt{P_z}$, which implies that the parameter
\begin{align}
\label{app-P1}
P_1
= \frac{\omega_1}{\omega_\text{rec,2}} \frac{1}{\omega_0^2t^2} P_z
\end{align}
is independent of $P_z$. Inserting $k_\pm= \bm e_\pm\cdot \bm k_2$ and Eq.\ \eqref{app-P1} into Eq.\ \eqref{app-theta-10-pm-eta} yields Eq.\ \eqref{theta-10-pm-cdot}.

Finally we note that if the pulse area is small, i.e.\ $\vartheta \ll 1$, then $P_{n',n}(\Delta_R)$ is well approximated by
\begin{align}
\label{app-P-vs-Delta-approx}
P_{n',n}(\Delta_R)
= \frac{\omega_{n',n}^2}{\Delta_R^2} \sin^2\frac{\Delta_R t}2
.\end{align}
The FWHM of this interaction-time broadened spectrum is easily found numerically, yielding $\Delta_{R,\text{FWHM}}\approx 2\pi \times 0.886/t$. One easily finds that a small pulse area implies
\begin{align}
\label{app-A-10-eta}
A_{n',n}
= \frac\pi2 \omega_{n',n}^2t
,&&
A_{1,0,\pm}
= \frac{\pi}{2} \omega_0^2 t \eta_\pm^2
.\end{align}

We now turn to a 1D situation and assume thermal equilibrium. Hence, the thermal population of the $n$th vibrational state is $p_n= (1-q)q^n$ with
\begin{align}
\label{app-q}
q
= e^{-\hbar\omega/k_BT}
.\end{align}
We also assume that the pulse area is small so that Eq.\ \eqref{app-A-10-eta} holds. Combining this with expressions analogous to Eq.\ \eqref{app-matrix-el-plus}, we obtain for the total areas under the blue and red sidebands
\begin{align}
A_\text{blue}
= \sum_{n=0}^\infty A_{n+1,n} p_n
= A_{1,0}\sum_{n=0}^\infty (1+n)p_n
= \frac{A_{1,0}}{1-q}
\end{align}
and
\begin{align}
\label{app-Ared-Ablue}
A_\text{red}
&= \sum_{n=1}^\infty A_{n-1,n} p_n
= A_{1,0}\sum_{n=1}^\infty n p_n
= A_{1,0}\frac{q}{1-q}
\notag \\ &
= q A_\text{blue}
.\end{align}
This yields $A_\text{red}/A_\text{blue}= q= e^{-\hbar\omega/k_BT}$ and $\overline n= q/(1-q)= (A_\text{blue}/A_\text{red}-1)^{-1}$. In addition, it yields $A_\text{blue}-A_\text{red}= A_{1,0}$ and combining this with Eq.\ \eqref{app-A-10-eta} yields the two-photon Rabi frequency in free space
\begin{align}
\label{app-omega-0}
\omega_0
= \frac1\eta \sqrt{2\frac{A_\text{blue}-A_\text{red}}{\pi t}}
.\end{align}

\end{document}